\documentclass{article} 
\usepackage{color}
\usepackage{amsmath,amsfonts,amsthm,amssymb}
\usepackage{eucal}
\usepackage[colorlinks=true,urlcolor=webblue,linkcolor=webgreen,filecolor=webblue,citecolor=webgreen,pdfpagemode=UseOutlines,pdfstartview=FitH,pdfpagelayout=OneColumn,bookmarks=true]{hyperref}
\usepackage{fullpage}

\hypersetup{
pdftitle=For Distinguishing Conjugate Hidden Subgroups the Pretty Good Measurement is as Good as it Gets,
pdfauthor=Cristopher Moore and Alexander Russell}

\definecolor{webgreen}{rgb}{0,.5,0}
\definecolor{webblue}{rgb}{0,0,.5}

\DeclareMathOperator{\Exp}{Exp}

\numberwithin{equation}{section}

\newtheorem{theorem}{Theorem}
\newtheorem{lemma}[theorem]{Lemma}

\newtheorem{claim}[theorem]{Claim}

\DeclareMathAlphabet{\varmathbb}{U}{bbold}{m}{n}
\newcommand{\one}{\varmathbb 1}
\newcommand{\Ind}{\textrm{Ind}}
\newcommand{\Res}{\textrm{Res}}
\newcommand{\planch}{\textrm{Planch}}

\renewcommand{\vec}[1]{\mathbf{#1}}

\newcommand{\success}{{\rm success}}

\newcommand{\remove}[1]{}

\newcommand{\C}{\mathbb{C}}
\newcommand{\R}{\mathbb{R}}
\newcommand{\Z}{\mathbb{Z}}

\newcommand{\ket}[1]{\left| #1 \right\rangle}
\newcommand{\bra}[1]{\left\langle #1 \right|}

\newcommand{\CG}{\C[G]}

\newcommand{\tr}{\textbf{tr}\,}

\newcommand{\rank}{\textbf{rk}\;}
\newcommand{\norm}[1]{\left\| #1 \right\|}

\newcommand{\inner}[2]{\left\langle #1 \!\mid\! #2 \right\rangle}

\newcommand{\vrho}{{\boldsymbol{\rho}}}
\newcommand{\vsigma}{{\boldsymbol{\sigma}}}

\newcommand{\Norm}{{\rm Norm}}
\newcommand{\wg}{\widehat{G}}
\newcommand{\bg}{\mathcal{B}}
\newcommand{\core}[2]{#1_{#2}}
\newcommand{\Reg}{R}
\newcommand{\mvsigma}{m}

\title{For Distinguishing Conjugate Hidden Subgroups, \\
the Pretty Good Measurement is as Good as it Gets}

\author{Cristopher Moore \\
\textsf{moore@cs.unm.edu}\\
Department of Computer Science\\
University of New Mexico
\and
Alexander Russell\\
\textsf{acr@cse.uconn.edu}\\
Department of Computer Science and Engineering\\
University of Connecticut}

\begin{document}
\maketitle 

\begin{abstract}
Recently Bacon, Childs and van Dam showed that the ``pretty good
measurement'' (PGM) is optimal for the Hidden Subgroup Problem on
the dihedral group $D_n$ in the case where the hidden subgroup is
chosen uniformly from the $n$ involutions.  We show that, for any
group and any subgroup $H$, the PGM is the optimal one-register
experiment in the case where the hidden subgroup is a uniformly random
conjugate of $H$. We go on to show that when $H$ forms a Gel'fand
pair with its parent group, the PGM is the optimal measurement for
any number of registers.  In both cases we bound the probability that the optimal 
measurement succeeds.  This generalizes the case of the dihedral group, 
and includes a number of other examples of interest.  
\end{abstract}

\section{The Hidden Conjugate Problem}

Consider the following special case of the Hidden Subgroup Problem,
called the \emph{Hidden Conjugate Problem} in~\cite{MooreRRS04}.  Let
$G$ be a group, and $H$ a non-normal subgroup of $G$; denote
conjugates of $H$ as $H^g = g^{-1} H g$.  Then we are promised that
the hidden subgroup is $H^g$ for some $g$, and our goal is to find out
which one.

The usual approach is to prepare a uniform superposition over the
group, entangle the group element with a second register by
calculating or querying the oracle function, and then measure the
oracle function.  This yields a uniform superposition over a random
left coset of the hidden subgroup,
\[ \ket{cH^g} = \frac{1}{\sqrt{|H|}} \sum_{h \in H^g} \ket{ch} \enspace . \]
Rather than viewing this as a pure state where $c$ is random, we may
treat this as a classical mixture over left cosets, giving the mixed
state with density matrix
\begin{equation}
\label{eq:rhog}
 \rho_g = \frac{1}{|G|} \sum_{c \in G} \ket{cH^g} \bra{cH^g} \enspace . 
\end{equation}
We then wish to find a positive operator-valued measurement (POVM) to
identify $g$.

A POVM consists of a set of positive \emph{measurement operators} $\{
E_i \}$ that obey the completeness condition
\begin{equation}
\label{eq:complete}
\sum_i E_i = \one \enspace . 
\end{equation}
If we are trying to distinguish $\ell$ density matrices $\{ \rho_i \mid 1 \leq
i \leq \ell \}$, and if $\rho_i$ is chosen with probability $p_i$, the
probability that the POVM gives the right answer is
\begin{equation}
\label{eq:success}
P_\success = \sum_i p_i \, \tr E_i \rho_i \enspace . 
\end{equation}

A theorem of Yuen, Kennedy and Max~\cite{YuenKM} and Holevo~\cite{Holevo} states that $P_\success$ is maximized if and only if the following conditions hold for every $i$:
\begin{gather}
 \left( \sum_j p_j \rho_j E_j - p_i \rho_i \right) E_i
 = 0\enspace, \label{eq:holevo}\\
\sum_j p_j \rho_j E_j  = \sum_j p_j E_j \rho_j\enspace,\; \text{and} \label{eq:commute} \\
\sum_j p_j \rho_j E_j  \geq p_i \rho_i\enspace, \label{eq:positive}
\end{gather}
where we write $A \geq B$ if $A-B$ is positive semidefinite.  
These conditions come from recognizing that maximizing $P_\success$ 
subject to the completeness condition~\eqref{eq:complete} gives a semidefinite program.  
Ip~\cite{Ip} used this fact to show that Shor's algorithm is optimal
for the hidden subgroup problem on the cyclic group $\Z_n$.

In a beautiful recent paper, Bacon, Childs and van Dam~\cite{BaconCvD}
consider the hidden conjugate problem for the dihedral group $D_n$
where $H$ is an order-2 subgroup; that is, when the hidden subgroup consists 
of the identity and one of the $n$ involutions.  They consider entangled
measurements over multiple registers each of which contains a coset
state, and show that the so-called ``pretty good
measurement''~\cite{PGM}, defined below, is in fact optimal for any
number of registers.  They then show that this optimal measurement is
related to random cases of the Subset Sum problem (see also
Regev~\cite{Regev}).  Other prior work includes Eldar, Megretski, and
Verghese~\cite{Eldar}, who showed that the PGM is optimal for certain
families of density matrices related by group symmetries; see references in that paper
for some other cases for which the PGM is optimal.

In this paper we point out that for any subgroup $H$ of a group $G$,
the pretty good measurement is the optimal one-register experiment for
finding a hidden conjugate of $H$ when all conjugates are equally
likely.  We then write a general expression for the probability
$P_\success$ that this measurement identifies the hidden conjugate in
a single experiment on a coset state. We then go on to show that when $G$ and
$H$ form a Gel'fand pair, the pretty good measurement is in fact the
optimal measurement on \emph{any number of registers}.  This recovers the optimality result 
of~\cite{BaconCvD} as a special case, and establishes
optimal measurements for a variety of other interesting group-subgroup
pairs, including the subgroups of the affine groups studied
in~\cite{MooreRRS04}.

We use the machinery of representation theory; we refer the reader
to~\cite{FultonH91,Serre77} or to the review in our
paper~\cite{MooreRS} for an introduction and for notation.

\section{The one-register case}

\subsection{The pretty good measurement is optimal}

The \emph{pretty good measurement} (PGM), also known as
the \emph{least squares measurement}, is defined as
follows~\cite{PGM}.  Given a set of density matrices $\rho_i$ with
associated probabilities $p_i$, let
\[ M = \sum_i p_i \rho_i . \]
Then the PGM associated with this family of mixed states is $\{ E_i \}$,
where the measurement operator $E_i$ is defined as
\begin{equation}
\label{eq:pgm}
E_i = p_i M^{-1/2} \rho_i M^{-1/2} 
\end{equation}
where the inverse $M^{-1/2}$ is defined on the image of $M$; that is,
$M^{-1/2}$ is the unique positive operator such that $(M^{-1/2})^2 M$ 
is the projection operator $\Pi$ onto the image of $M$.  
If $M$ has full rank, it is easy to see that this choice of 
$\{ E_i \}$ satisfies the completeness condition~\eqref{eq:complete}; if the image 
of $M$ is a proper subspace, we satisfy the completeness condition by adding
an additional measurement which projects onto its orthogonal complement.

We will show that the optimality
conditions~\eqref{eq:holevo},~\eqref{eq:commute},
and~\eqref{eq:positive} hold for the PGM for the family of density
matrices $\{ \rho_g \}$ defined above when endowed with the uniform
distribution $p_g = 1/|G|$.  First we derive the structure of the PGM.
Observe that the mixed state $\rho_g = |G|^{-1} \sum_c \ket{cH^g}\bra{cH^g}$
has the property that
\begin{equation}
\label{eq:rhosquare}
\rho_g^2 = \frac{|H|}{|G|} \rho_g
\end{equation}
so that $\rho_g$ is a projection operator scaled by the constant
$|H|/|G|$ (this follows from the fact that the uniform distribution on 
any subgroup is its own square under convolution).  
As $\tr \rho_g = 1$, we must have $\rank \rho_g = |G|/|H|$, 
the index of $H$ in $G$.

Note that $\rho_g$ commutes with the left action of $G$, since it is ``symmetrized'' 
over all left cosets.  By Schur's lemma, $\rho_g$ is block diagonal, with blocks corresponding 
to the decomposition of $\CG$ into bi-invariant spaces. Furthermore, the
block corresponding to the irreducible representation $\sigma$ has form
$A_\sigma \otimes \one_{d_\sigma}$, where $\one_{d_\sigma}$ acts within each
left-invariant space.  Recalling~\eqref{eq:rhosquare}, $A_\sigma$ is a rescaled
projection operator, and we may write $\rho_g = \oplus_{\sigma \in \wg} \,\rho_g^\sigma$ where
\begin{equation}
\label{eq:rhogsigma}
\rho_g^\sigma = \frac{|H|}{|G|}  \pi_g^\sigma \otimes \one_{d_\sigma}\enspace;
\end{equation}
here $\pi_g^\sigma$ is the projection operator
\begin{equation}
\label{eq:pi}
 \pi_g^\sigma = \frac{1}{|H|} \sum_{h \in H^g} \sigma(h) \enspace . 
\end{equation}

Since $\rank \pi_g^\sigma$ does not depend on $g$, we denote it simply as
$\rank \pi^\sigma$.  Then the following lemma describes the structure of
the PGM for $\{ \rho_g \}$.  Since $\rho_g = \rho_h$ whenever $g$ and $h$
are in the same coset of the normalizer $\Norm(H) = \{ g \mid H^g = H
\}$, without loss of generality we assume that the POVM gives a
uniformly random element of some coset of $\Norm(H)$, and we count it
as having succeeded if it gives an element of the correct coset.  Thus
we will multiply $P_\success$ by the index of $\Norm(H)$ below.
 
\begin{lemma}
\label{lem:pgm}
For the family $\{ \rho_g \}$ of density matrices corresponding to random left cosets of conjugate subgroups $H^g$ with the uniform distribution on $g$, the pretty good measurement operators $E_g$ are given by
\[ E_g = \bigoplus_{\sigma \in \wg} E_g^\sigma \]
and, for each $\sigma \in \wg$,
\begin{equation}
\label{eq:eg}
 E_g^\sigma = \frac{d_\sigma}{|G| \rank \pi^\sigma} \pi_g^\sigma \otimes \one_{d_\sigma}
\end{equation}
where $\pi_g^\sigma$ is defined as in~\eqref{eq:rhogsigma}.
\end{lemma}

\begin{proof}  
For each $x \in G$, let $L_x$ and $R_x$ denote the unitary operators
that carry out left and right group multiplication by $x$.  Note that left cosets are mapped to each other by $L_x$ and, in particular, $\ket{cH} = L_c \ket{H}$.  Note furthermore that $R_x$ maps left cosets of one conjugate onto left cosets of another conjugate, e.g., $\ket{H^g} = R_g \ket{g^{-1} H}$, and that $L_x$ commutes with $R_y$ for all $x,y$.

We have $p_g = 1/|G|$ for all $g$.  Now if we write
\begin{align*}
 M &= \frac{1}{|G|} \sum_g \rho_g = \frac{1}{|G|^2} \sum_{c,g} \ket{cH^g} \bra{cH^g} 
 = \frac{1}{|G|^2} \sum_{c,g} \ket{cHg} \bra{cHg} \\
&= \frac{1}{|G|^2} \sum_{c,g} L_c R_g \ket{H} \bra{H} R_g^\dagger L_c^\dagger
\end{align*}
(where in the third equality we replace $cg^{-1}$ with $c$), 
we see that $M$ commutes with $L_x$ and $R_x$ for all $x \in G$.  That is, summing over both the left coset and the choice of conjugate ``symmetrizes'' $M$ on both the left and the right.  It follows by Schur's lemma that $M$ takes the form 
\[ M = \bigoplus_{\sigma \in \wg} M^\sigma \]
where $M^\sigma$ is a scalar multiple of the identity operator for each $\sigma$. As
\[ M_\sigma = \frac{1}{|G|} \sum_g \rho_g^\sigma = \frac{|H|}{|G|^2} \sum_g \pi_g^\sigma \otimes \one_{d_\sigma} \enspace , \]
by taking traces we conclude that
$$
M^\sigma = \left( \frac{|H|}{|G|} \frac{\rank \pi^\sigma}{d_\sigma}\right) \one_{d_\sigma^2}
\enspace . 
$$

Similarly, $\rho_g$ is block-diagonal, as it commutes with $L_x$
(though not with $R_x$ unless $H$ is normal).  Therefore, $M$ commutes with
$\rho_g$ for each $g$, and~\eqref{eq:pgm} becomes
\begin{equation}
\label{eq:infact}
E_g = \frac{1}{|G|} M^{-1} \rho_g 
\end{equation}
giving, in each irreducible block, 
\begin{equation}
\label{eq:pgm-operators-single}
E_g^\sigma = \frac{d_\sigma}{|H|\, \rank \pi^\sigma} \rho_g^\sigma 
= \frac{d_\sigma}{|G| \,\rank \pi^\sigma} \pi_g^\sigma \otimes \one_{d_\sigma} 
\end{equation}
which completes the proof.
\end{proof}

We now give our proof that the PGM is optimal for the hidden conjugate problem for any $G$ and $H$.  This follows simply from the fact that $M$, and therefore $E_g$, commutes with $\rho_g$ for each $g$.

\begin{theorem}
\label{thm:pgm}  
For the family $\{ \rho_g \}$ of density matrices corresponding to random left cosets of conjugate subgroups $H^g$ with the uniform distribution on $g$, the pretty good measurement $\{ E_g \}$ optimizes the probability of correctly measuring $g$.
\end{theorem}

\begin{proof}
  Since the $E_g$ and $\rho_g$ are block-diagonal (according to the same
  decomposition of $\CG$), it suffices to confirm the optimality
  criteria~\eqref{eq:holevo}, \eqref{eq:commute},
  and~\eqref{eq:positive} in each block, i.e., for each $\sigma \in \wg$.

First,~\eqref{eq:commute} holds trivially since $E_g^\sigma$ is proportional to $\rho_g^\sigma$ 
for each $g$, and so commutes with it.
As for condition~\eqref{eq:positive}, observe that from~\eqref{eq:rhosquare} 
and~\eqref{eq:pgm-operators-single} we have
\begin{equation}
\label{eq:rhogeg}
 \rho_g^\sigma E_g^\sigma = \frac{d_\sigma}{|G| \,\rank \pi^\sigma} \rho_g^\sigma 
\end{equation}
and so for any $h \in G$, recalling that $p_g = 1/|G|$ for all $g$, we have
\begin{align*}
\frac{1}{|G|} \sum_g \rho_g^\sigma E_g^\sigma 
&= \frac{d_\sigma}{|G| \,\rank \pi^\sigma} M^\sigma
= \frac{|H|}{|G|^2} \left( \one_{d_\sigma} \otimes \one_{d_\sigma} \right) 
\geq \frac{|H|}{|G|^2} \left( \pi_h^\sigma \otimes \one_{d_\sigma} \right)
= \frac{1}{|G|} \rho_h
\end{align*}
since $\one_{d_\sigma} \geq \pi_h^\sigma$.  Finally, for any $h \in G$
  \begin{align*}
  \left( \sum_g \rho_g^\sigma E_g^\sigma - \rho_h^\sigma \right) E_h^\sigma
  &= \frac{|H|}{|G|} \left[ \left( \one_{d_\sigma} - \pi_h^\sigma \right) \otimes \one_{d_\sigma} \right] E_h^\sigma 
  = \frac{|H| \,d_\sigma}{|G|^2 \,\rank \pi^\sigma} \left[ \left( \one_{d_\sigma} - \pi_h^\sigma\right) \pi_h^\sigma \right]
  \otimes \one_{d_\sigma} = 0
\end{align*}
since $(\one-\pi)\pi = 0$ for any projection operator $\pi$.  Again 
recalling $p_g = 1/|G|$ establishes~\eqref{eq:holevo} and completes the proof.
\end{proof}
 
Note that Lemma~\ref{lem:pgm} and Theorem~\ref{thm:pgm} imply that, as pointed out before~\cite{Ip,Kuperberg03,MooreRS}, the optimal measurement consists of first measuring the representation name $\sigma$, and then performing an additional measurement $M_g^\sigma$ inside the bi-invariant space corresponding to $\sigma$.

\subsection{Partial measurements}

Suppose that rather than trying to identify the conjugate exactly, we wish to learn some partial information about it.  To learn one bit, for instance, we would divide the set of conjugates into two equal subsets, and combine the $\rho_g$ into two mixed states $\rho_0$ and $\rho_1$ consisting of mixtures of those in the two subsets.  The next theorem shows that the PGM is optimal for any such partial measurement as long as each subset of the set of conjugates has probability proportional to its size.  

\begin{theorem}
\label{thm:partial}
Let $C(H)=\{ H^g \}$ be partitioned into disjoint sets $C_i$, $1 \leq i
\leq \ell$.  Let 
\[ \rho_i = \frac{1}{|C_i|} \sum_{g: H^g \in C_i} \rho_g \]
where $\rho_g$ is as in~\eqref{eq:rhog}, and let $p_i = |C_i|/|C(H)|$.
Then the pretty good measurement for the family $\{ \rho_i \}$ with
probabilities $p_i$ is the family of operators $\{ E_i \}$
\[ E_i = \sum_{g : H^g \in C^i} E_g \enspace , \]
where $E_g$ is given by Lemma~\ref{lem:pgm};
 this measurement is optimal.
\end{theorem}

\begin{proof}  The proof is exactly the same as that of Lemma~\ref{lem:pgm} and Theorem~\ref{thm:pgm}, except that for each $i$ we sum over the $g$ with $H^g \in C_i$.
\end{proof}

\subsection{The probability of success}

As a corollary to Theorem~\ref{thm:pgm}, we can determine the optimal
success probability. Let $C(H)$ denote the set of conjugates of $H$. 
For a group $G$ the \emph{Plancherel measure} is the probability distribution on $\wg$
assigning $\sigma \in \wg$ the probability $d_\sigma^2/|G|$; this is the fraction, dimensionwise, 
of $\CG$ consisting of the bi-invariant subspace corresponding to $\sigma$.  
For a set $S \subset \wg$, we let $\planch(S)$ denote the probability of
observing an element of $S$ according to the Plancherel measure. We
remark that the Plancherel measure is precisely the probability
distribution obtained by performing weak Fourier sampling when the 
hidden subgroup is trivial, since in that case the state $\rho$ is completely mixed.

Given a subgroup $H$ of $G$, let $S_H \subseteq \wg$ denote the 
set of irreducible representations for which the projection operator 
$\pi_H^\sigma = |H|^{-1} \sum_{h \in H} \sigma(h)$ is nonzero, 
and let $\core{H}{G} = \cap_g H^g$ denote the largest normal subgroup
contained in $H$.  Then the following theorem gives the success probability of the 
optimal one-register experiment.  
\begin{theorem}
\label{thm:pgm-success-single}
Given a group $G$ and a subgroup $H$, the probability that
the optimal single-register measurement correctly identifies a
uniformly random conjugate of $H$ is
\begin{equation}
  \label{eq:psuccess-single}
  P_\success = \frac{|H|}{|C(H)|} \planch(S_H) \leq \frac{|H|}{|C(H)| \cdot |\core{H}{G}|} \enspace .
\end{equation}
\end{theorem}

\begin{proof}
Let us say that $g \sim g'$ if $H^g = H^{g'}$ (and so $\rho_g = \rho_{g'}$).  
Then $P_\success$ is the expectation 
\[ P_\success = \Exp_g \sum_{g' \sim g} \tr E_{g'} \rho_g \enspace , \]
where $g$ is selected uniformly in $G$.  Since $g \sim g'$ if and only if 
$g$ and $g'$ are in the same (right) coset of the normalizer 
$\Norm(H) = \{ g \mid H^g = H \}$, we have
\[
P_\success = |\Norm(H)| \cdot \Exp_g \tr E_{g} \rho_g 
= |\Norm(H)| \cdot \Exp_g \sum_\sigma \tr E_{g}^\sigma \rho_g^\sigma \enspace .
\]
Considering~\eqref{eq:rhogsigma} and~\eqref{eq:rhogeg}, we have
\[ \tr E_g^\sigma \rho_g^\sigma = \frac{|H| \,d_\sigma^2}{|G|^2}
= \frac{|H|}{|G|} \planch(\sigma) \enspace , \] 
but only for those $\sigma$ where $\rho_g^\sigma$ and $E_g^\sigma$ are nonzero, 
i.e., those for which $\rank \pi^\sigma > 0$.  Thus, we conclude that
\[
P_\success = \frac{|\Norm(H)| |H|}{|G|} \cdot \sum_{\sigma \in S_H} \planch(\sigma) 
= \frac{|H|}{|C(H)|} \cdot\planch(S_H) 
\]
where we recall that $|C(H)| = |G|/|\Norm(H)|$.

Now, note that for any subgroup $K \subseteq H$, we have $S_H \subseteq S_K$ 
since any $\sigma$ that annihilates $K$ also annihilates $H$.  
In addition, if $K$ is normal, recall that for any $\sigma$ 
we have either $\pi_K^\sigma = 0$ or $\pi_K^\sigma = \one$.  It follows that 
$\sum_{\sigma \in S_K} d_\sigma^2 = \sum_{\sigma \in S_K} d_\sigma \rank \pi_K^\sigma 
= \rank \pi_K^\Reg$ where $\Reg$ is the regular representation, and since 
$\rank \pi_K^\Reg = |G|/|K|$ we have $\planch(S_K) = 1/|K|$.  
Thus $\planch(S_H) \leq 1/|\core{H}{G}|$, completing the proof of~\eqref{eq:psuccess-single}.
\remove{
We remark, however, that $\rank \pi^\sigma > 0$ exactly when $\langle \Res_{H}
\chi_{\sigma}, 1\rangle_H > 0$ and hence, by Frobenius reciprocity
(see~\cite{Serre77}), that $\langle \chi_{\sigma}, \Ind_{H} 1\rangle_G > 0$.
Specifically, $S$ consists of precisely those representations that
occur in the representation $\Ind_{H} 1$. It is easy to argue that
$\ker (\Ind_H^G 1) = \core{H}{G} =
\cap_{\ell \in L} H^\ell$; hence $S \subseteq \{ \sigma \mid \core{H}{L} < \ker \sigma\}$
and $\sum_{\sigma \in S} d_\sigma^2 \leq |G| / |\core{H}{G}|$. The
inequality of~\eqref{eq:psuccess-single} follows.
}
\end{proof}

Note that if we observe any $\sigma \notin S_H$, we know that the promise 
that the hidden subgroup is a conjugate of $H$ has been violated.  Thus, 
as in~\cite{BaconCvD}, if we are promised that the hidden subgroup is
either trivial or a conjugate of $H$, we can complete the PGM with an
additional measurement operator $M_0$ that projects onto the orthogonal 
complement of $S_H$, and conclude that the hidden subgroup is trivial if we observe
the outcome $M_0$.

It is interesting to compare Theorems~\ref{thm:pgm} and~\ref{thm:pgm-success-single} 
with known results on the hidden subgroup
problem.  For the dihedral group $D_n$ where $H$ is an order-2
subgroup, there are $n$ conjugates, and $S_H$ consists of all of $\wg$ except 
for the sign representation.  Thus we have
\[ P_\success = \frac{2}{n} \left( 1- \frac{1}{2n} \right) \enspace . \]
On the other hand, for the affine group $A_p$, the maximal subgroup $H=\Z_p^*$ has 
$p$ conjugates, and $S_H$ includes all but the $p-2$ nontrivial one-dimensional 
representations.  This gives
\[ P_\success = \frac{p-1}{p} \left( 1-\frac{p-2}{p(p-1)} \right) = 1-\frac{2(p-1)}{p^2} \enspace . \]
Indeed, Moore, Rockmore, Russell and Schulman~\cite{MooreRRS04} gave an explicit algorithm using a von Neumann measurement that succeeds with constant probability.  This algorithm can easily be modified to carry out the optimal POVM in polynomial time; see also Bacon, Childs and van Dam~\cite{BaconCvDHeisenberg}.

Now let us consider the case of the hidden subgroup problem relevant to Graph Isomorphism in the case of two rigid, connected graphs of size $n/2$.  Here $G=S_n$ and $H$ is the order-2 subgroup consisting of $n/2$ disjoint transpositions,
\[ H = \{1, (1\,2)(3\,4)\cdots(n-1\,n) \} \]  
of which there are $(n-1)!!$ conjugates, one for each perfect matching of $n$ items.  
Using lemmas proved in~\cite{MooreRS}, it is easy to show that for almost all representations $\sigma$ 
(with respect to the Plancherel distribution) we have $\rank \pi^\sigma =  (1 \pm o(1)) d_\sigma / 2$, 
so $\planch(S_H) = 1-o(1)$.  Thus we have
\[ P_\success = \frac{2}{(n-1)!!} (1-o(1)) = n^{-n/2} e^{O(n)} \]
This can be generalized to other conjugacy classes using general character bounds due to Roichman~\cite{Roichman96}; see also Kempe and Shalev~\cite{KempeS}.

However, it should be emphasized that the fact that $P_\success$ is exponentially small does not mean that we need an exponential number of single-register experiments to solve the hidden conjugate problem.  In particular, Ettinger and H{\o}yer~\cite{EttingerH98} showed that a polynomial number (i.e., $O(\log |G|) = O(\log n)$) of single-register experiments is enough to determine, information-theoretically, an involution in $D_n$.  Thus our results here do not subsume the results of  Moore, Russell and Schulman~\cite{MooreRS} and Moore and Russell~\cite{MooreR}, who showed that it takes an exponential number of single-register experiments,  or a super-polynomial number of two-register experiments, to obtain \emph{even a single bit of information} about the conjugate of $H$ in $S_n$.

\section{Multiregister measurements and Gel'fand pairs}

For the multiregister experiment, we view states as elements of the
Hilbert space $\C[G^k] = \CG^{\otimes k}$.  We now have a random left coset of the subgroup 
$H^k \subset G^k$, and the corresponding mixed state is
$$
\vrho_g = \rho_g^{\otimes k} = \frac{1}{|G|^k} \sum_{\vec{c} \in G^k} \ket{ \vec{c} (H^g)^k}\bra{ \vec{c} (H^g)^k}\enspace.
$$
Since $\rho_g^{\otimes k}$ is symmetrized over left cosets, it commutes with 
left multiplication in $G^k$.  Thus by Schur's lemma it is block-diagonal,
where each block corresponds to 
a representation $\vsigma = \sigma_1 \otimes \cdots \otimes \sigma_k$ 
of $G^k$, and each $\sigma_i$ is an irreducible representation of $G$.  
Indeed, in a given such block we can write
\begin{equation}
\label{eq:vrho}
 \vrho_g^{\vsigma}
= \rho_g^{\sigma_1} \otimes \cdots \otimes \rho_g^{\sigma_k} 
\enspace . 
\end{equation}

The situation in the multiregister case is complicated by the fact that, unlike the one-register case, $M$ does not generally commute with $\vrho_g$.  Indeed, they do not commute even for the two-register case in the dihedral group.  We note in passing that they do commute in a few special cases: for instance, when $H$ is generated by an involution that commutes with its conjugates.  However, this is not a very interesting case, since then $H$ and its conjugates generate an Abelian subgroup $K \subset G$, and we can distinguish them by solving the hidden subgroup problem on $K$.

However, we can still prove that the PGM is optimal in the case that $G$ and $H$ form a Gel'fand pair; we review the definition here, and also refer the reader to~\cite{Terras99} for an introduction.
Given a group $G$ and a subgroup $H$, let $\bg =
\bg_H(G)$ denote the collection of functions $f: G \to \C$
that are invariant under both left and right multiplication by $H$, 
i.e., such that $f(hg) = f(g) = f(gh)$ for all $g \in G$ and $h \in H$.
This collection of \emph{bi-invariant} functions forms a natural
algebra under convolution, and $\bg$ can be identified
with the subalgebra of $\CG$ generated by elements corresponding
to double cosets,
$HgH = (\sum_{h \in H} h) \cdot g \cdot (\sum_{h \in H} h)$.
Then the following criteria are equivalent, and the pair $(G,H)$ 
is said to be \emph{Gel'fand} if any of them hold: 
\begin{enumerate}
\item $\bg$ is commutative. 
\item The induced representation $\Ind_H^G \one$ contains no more than
one copy of any particular $\sigma \in \wg$.
\item For any $\sigma \in \wg$ and any $f \in \bg$, the Fourier transform $\hat{f}(\sigma) = \sum_g f(g) \sigma(g)$ has rank at most one.
\end{enumerate}
The third criterion is the one most relevant to our analysis.  
Suppose $(G,H)$ is a Gel'fand pair; then since the uniform distribution on $H$ is an element of $\bg$, 
for any $\sigma \in \wg$ the projection operator $\pi^\sigma = |H|^{-1} \sum_{h \in H} \sigma(h)$
has rank at most one.  Since this is also true of its conjugates $\pi_g^\sigma = \sigma(g)^{-1} \pi^\sigma \sigma(g)$, 
we see that $(G,H^g)$ is Gel'fand for all $g \in G$.

As stated above, Bacon, Childs and van Dam~\cite{BaconCvD} showed that the pretty good measurement is optimal for the dihedral groups $D_n$ when the hidden subgroup $H$ is of order 2.  Indeed, $(D_n,H)$ is Gel'fand for these subgroups, and we generalize their result as follows.

\begin{theorem}

\label{thm:gelfand}  
For any number of registers $k > 0$, given the family $\{ \vrho_g \}$ of density matrices corresponding to random left cosets of conjugate subgroups $(H^g)^k \subset G^k$ with the uniform distribution on $g$, the pretty good measurement $\{ E_g \}$ optimizes the probability of correctly measuring $g$.
\end{theorem}

\begin{proof} As before, we will show that the optimality conditions
hold in each irreducible block, since the $\vrho_g$, and therefore
$M$ and the $E_g$, are block-diagonal.  Since the tensor product of
rank-one operators has rank one, given $\vsigma = \sigma_1 \otimes \cdots \otimes \sigma_k$
with $\sigma_i \in \wg$ for all $i$, from~\eqref{eq:vrho} we have either
$\vrho_g^{\vsigma} = 0$ or
\[ \vrho_g^{\vsigma} = \ket{v_g} \bra{v_g} \otimes \one_{d_\vsigma} \]
for some vector $v_g \in \vsigma$.  In the latter case we have
$M^\vsigma = \mvsigma \otimes \one_{d_\vsigma}$ where
\[ \mvsigma = \frac{1}{|G|} \left( \sum_g \ket{v_g} \bra{v_g} \right)\]
and 
\[ \vrho_g^{\vsigma} E_g^\vsigma 
= \frac{1}{|G|} \left(\ket{v_g} \bra{v_g} \mvsigma^{-1/2} \ket{v_g} \bra{v_g} \mvsigma^{-1/2}
\right) \otimes \one_{d_\vsigma} = \frac{C}{|G|} \vrho_g^{\vsigma}
(M^\vsigma)^{-1/2}
\]
where $C$ is defined as the inner product
\[ C = \bra{v_g} \mvsigma^{-1/2} \ket{v_g} \enspace . \]
Similarly, $E_g^\vsigma \vrho_g^{\vsigma} = \frac{C}{|G|} 
(M^\vsigma)^{-1/2} \vrho_g^{\vsigma}$.

It is easy to see that $C$ does not depend on $g$: if $R_g$ denotes
the unitary operator corresponding to right multiplication by the
diagonal element $(g,\ldots,g)$ in $\C[G^k]$, then $\ket{v_g} = R_g
\ket{v_1}$ and $\mvsigma$ commutes with $R_g$.  Thus $C = \bra{v_g}
\mvsigma^{-1/2} \ket{v_g} = \bra{v_1} R_g \mvsigma^{-1/2} R_g^\dagger \ket{v_1} =
\bra{v_1} \mvsigma^{-1/2} \ket{v_1}$.  Then we have
\[ \sum_g \vrho_g^{\vsigma} E_g^\vsigma 
= \frac{C}{|G|} \sum_g \vrho_g^{\vsigma} (M^\vsigma)^{-1/2}
= C (M^\vsigma)^{1/2} 
= \frac{C}{|G|} \sum_g (M^\vsigma)^{-1/2} \vrho_g^{\vsigma} 
= \sum_g E_g^\vsigma \vrho_g^{\vsigma} 
\]
confirming~\eqref{eq:commute}.

As for condition~\eqref{eq:positive}, recalling the equality $\sum_g
\vrho_g^{\vsigma} E_g^\vsigma = C (M^\vsigma)^{1/2}$ just above, 
\eqref{eq:positive} is equivalent to the condition 
$C (M^{\vsigma})^{1/2} \geq \vrho_g^{\vsigma}$, and hence to
$C \mvsigma^{1/2} \geq \ket{v_g} \bra{v_g}$.  
Furthermore, this holds if and only if 
\[ \bra{v_g} C \mvsigma^{1/2} \ket{v_g} 
\geq \inner{v_g}{v_g} \inner{v_g}{v_g} 
= \norm{v_g}^4 \]
or, expanding the definition of $C$, 
\begin{equation}
\label{eq:positivity-gelfand}
\bra{v_g}\mvsigma^{1/2}\ket{v_g}\bra{v_g}\mvsigma^{-1/2}\ket{v_g}
\geq \norm{v_g}^4 \enspace . 
\end{equation}

We remark that if a positive semidefinite operator $A$ is a linear combination 
$A = \sum_\alpha a_\alpha B_\alpha$ of positive semidefinite operators $B_\alpha$
with $a_\alpha \in \R^+$, then the kernel of $A$ is $\bigcap_\alpha \ker B_\alpha$. 
As $\mvsigma$ is such a linear combination of operators $\ket{v_g}\bra{v_g}$,
it follows that $\ket{v_g}$ is orthogonal to $\ker \mvsigma$, and therefore lies in the 
image of $\mvsigma$.  Thus we can regard $\mvsigma^{1/2}$ and $\mvsigma^{-1/2}$ 
as inverses, and the inequality~\eqref{eq:positivity-gelfand} follows from the
following claim:

\begin{claim}
Let $A$ be a positive operator on a finite dimensional Hilbert space
$V$ and let $\ket{v} \in V$. Then
\[
\bra{v} A \ket{v} \bra{v} A^{-1} \ket{v} \geq 
\norm{v}^4 \enspace .
\]
\end{claim}

\begin{proof}
For a vector $w = (w_1, \ldots, w_n)$ of non-negative weights and a real
number $r$, define
\[
\mathcal{M}_r^w(x_1, \ldots, x_n) = \left(\frac{\sum_i w_i x_i^r}{\sum_i
    w_i}\right)^{1/r}
\]
to be the \emph{weighted $r$-mean} of the positive vector $x = (x_1,
\ldots, x_n)$. The \emph{power mean inequality}
(cf.~\cite[\S2.9]{HardyLP}) asserts that for $r < s$ we have
$\mathcal{M}_r^w(x) \leq \mathcal{M}_s^w(x)$.

The claim follows immediately from the from power mean inequality
with $r = -1$ and $s = 1$. Specifically, let $B = \{ \ket{b_i} \}$
be a spectral resolution of $A$, so that $B$ is an orthogonal basis
of eigenvectors for $V$, and let
$\lambda_i > 0$ be the associated eigenvalues, so that $\lambda_i \ket{b_i} =
A \ket{b_i}$. Writing $\ket{v} = \sum_i v_i \ket{b_i}$, we have
\[
\bra{v} A \ket{v} = \sum_i |v_i|^2 \lambda_i\qquad\text{and}\qquad\bra{v}
A^{-1} \ket{v} = \sum_i |v_i|^{2} \lambda_i^{-1}\enspace.
\]
If we adopt the weights $w_i = |v_i|^2$, then $\sum_i w_i = \norm{v}^2$ 
and the claim is equivalent to the the power mean
inequality $\mathcal{M}_{-1}^w(\lambda_1, \ldots, \lambda_n) \leq
\mathcal{M}_{1}^w(\lambda_1, \ldots, \lambda_n)$.  
\end{proof}

Returning to the proof of Theorem~\ref{thm:gelfand}, it remains to
establish~\eqref{eq:holevo}. Consider
 \begin{align}
 \left( \sum_g \vrho_g^{\vsigma} E_g^\vsigma - \vrho_h^{\vsigma} \right) E_h^\vsigma
 &= \frac{1}{|G|} \left[\left( C \mvsigma^{1/2} - \ket{v_h} \bra{v_h} \right) 
   \mvsigma^{-1/2} \ket{v_h} \bra{v_h} \mvsigma^{-1/2}\right] \otimes \one_{d_\vsigma} \nonumber \\
 &= \frac{C}{|G|} \left[ \left( \Pi - \one \right) \ket{v_h} \bra{v_h}
   \mvsigma^{-1/2} \right] \otimes \one_{d_\vsigma}
 \label{eq:why}
 \end{align}
where $\Pi$ is the projection operator onto the image of $\mvsigma$.  Since $\ket{v_h}$ 
lies in this image as discussed above, we have $\Pi \ket{v_h} = \ket{v_h}$ and~\eqref{eq:why} 
is identically zero.  
\end{proof}

As in the one-register case, we can distinguish the trivial subgroup from the conjugates of $H$ by completing the PGM with a measurement $M_0$ that projects onto the complement of the image of $M$.

\subsection{The Probability of Success}

In this section we give an upper bound on the success probability of the 
optimal multiregister experiment for Gel'fand pairs.  To prepare for this, 
we record a version of Holevo's theorem on
the capacity of a quantum channel~\cite{Holevo73b}.
\begin{lemma}
Let $R = \{ \rho_i \mid i \in I\}$ be a family of density matrices treated as
linear operators on the Hilbert space $V$. Let $E = \{ E_i \}$ be a
family of measurement operators on $V$ for which $\sum_i E_i = \one$
and, for each $i \in I$, $E_i$ is a scalar multiple of a projection
operator of rank $r$. Then
$$
\Exp_i \tr E_i \rho_i \leq \frac{\dim H}{r |I|}\enspace,
$$
where the index $i$ is chosen uniformly at random in $I$.
\end{lemma}

\begin{proof}
By assumption, we may write $E_i = \alpha_i \Pi_i$ where $\Pi_i$ is a
projection operator of rank $r$ and $\alpha_i \geq 0$. As $\rho_i$ is a density
matrix, we may write $\rho_i = \sum_j p_j \ket{v_j}\bra{v_j}$ where each
$\ket{v_j}$ has unit length, each $p_j \in [0,1]$, and $\sum_j p_j = 1$.
Observe that
$$
\tr E_i \rho_i = \sum_j p_j \bra{v_j} E_i \ket{v_j} \leq \sum_j p_j \|E_i\| = \|
E_i \| = \alpha_i\enspace,
$$
where $\| A \|$ is the operator norm of $A$, given by $\| A \| =
\max_{\vec{v} \neq 0} \| A\vec{v} \|/ \| \vec{v} \| $. Hence
$$
\Exp_i \tr E_i \rho_i \leq \frac{\sum_i \alpha_i}{|I|}\enspace.
$$
Observe now that $\sum_i E_i = \one$ and hence that $\dim V = \tr \one
= \tr \sum_i E_i = \sum_i \alpha_i r$; evidently $\sum_i \alpha_i = \dim V / r$, which
completes the proof.
\end{proof}

\begin{theorem}
\label{thm:pgm-success-gelfand}
Let $H$ be a subgroup of $G$ for which $(G,H)$ is a Gel'fand pair
and let $\core{H}{G} = \cap_g H^g$ be the largest normal subgroup
contained in $H$. The probability that the optimal $k$-register
measurement correctly identifies a uniformly random conjugate of $H$
is
\begin{equation}
  \label{eq:pgm-success-gelfand}
  P_\success \leq \frac{1}{|C(H)|} \cdot \left(\frac{|H|}{|\core{H}{G}|} \right)^k
\end{equation}
\end{theorem}

\begin{proof}
Without sacrificing optimality, we may initially carry out weak
Fourier sampling, in which we observe a representation $\vsigma$
with probability
\[
P(\sigma) = \frac{|H|^k d_\vsigma \,\rank \pi_g^\vsigma}{|G|^k} \enspace .
\]
Again accounting for the fact that $\vrho_g = \vrho_{g'}$ when $H^g =
H^{g'}$, the probability of success may then be written 
\[
P_\success =
|\Norm(H)| \cdot \Exp_g \tr E_g \vrho_g = |\Norm(H)| \cdot \Exp_g
\Exp_{\vsigma} a_\vsigma \tr E_g^\vsigma \vrho_g^\vsigma
\]
where $\vsigma$ is distributed according to $P(\vsigma)$, $g$ is
uniform in $G$, and $a_\vsigma = 1/\tr \vrho^\sigma_g$ normalizes
$\vrho_g^\vsigma$ so that it is a density matrix.

Recall that $\vrho_g^\vsigma$ is proportional to $\pi_g^\vsigma \otimes \one_{d_\vsigma}$ 
where $\pi_g^\vsigma$ has rank one (or zero), that $M^\vsigma = m \otimes \one_{d_\vsigma}$, 
and that the image of $\pi_g^\vsigma$ is contained in the image of $m$.
Therefore, $m^{-1/2} \pi_g^\vsigma m^{-1/2}$ has rank one or zero, and so 
$E_g^\vsigma = m^{-1/2} \pi_g^\vsigma m^{-1/2} \otimes \one_{d_\vsigma}$ 
is either zero or a scalar multiple of a projection operator of rank $d_\vsigma$.  
By the lemma above, for a representation $\vsigma$ we have $\Exp_g \tr(a_\vsigma
E_g^{\vsigma} \vrho^\vsigma_g) \leq d_\vsigma / |G|$. 

Let us again define $S_{H^k} \subset \widehat{G^k}$ to be the set
of representations for which $\pi^\vsigma_g$ is nonzero.  
By commuting the two expectations above, we conclude that
\[
P_\success \leq |\Norm(H)| \Exp_\vsigma \frac{d_\vsigma}{|G|} 
= \frac{|\Norm(H)|}{|G|} |H|^k \sum_{\vsigma} \rank \pi^\vsigma \frac{d_\vsigma^2}{|G|^k} 
= \frac{|H|^k}{|C(H)|} \planch(S_{H^k})
\]
where $\planch_{G^k}(\vsigma) = d_\vsigma^2/|G|^k$ is the Plancherel distribution on 
$\widehat{G^k}$.  But this is just the product of the Plancherel distribution on $\wg$ over the 
$\sigma_i$, so we have
\[ 
P_\success \leq \frac{|H|^k}{|C(H)|} \planch(S_H)^k
\]
and recalling from the proof of Theorem~\ref{thm:pgm-success-single} that 
$\planch(S_H) \leq 1/|H_G|$ completes the proof.
\end{proof}

\subsection{Examples}

Theorem~\ref{thm:gelfand} applies to a number of group families that have appeared in the literature on the hidden subgroup problem.  Here is a short list of examples of Gel'fand pairs:
\begin{itemize}
\item $(G,H)$ where $H$ is normal and $G/H$ is Abelian.  Of course, whenever $H$ is normal the hidden conjugate problem becomes trivial.
\item $(D_n,H)$ where $H$ consists of the identity and an involution, as in Bacon, Childs and van Dam~\cite{BaconCvD}.
\item $(A_p,\Z_p^*)$ where $A_p$ is the affine group $\Z_p^* \ltimes
\Z_p$ and $\Z_p^*$ is a maximal non-normal subgroup.  An efficient
quantum algorithm for the hidden conjugate problem in this case was given by~\cite{MooreRRS04}.
\item All the subgroups of the Heisenberg group, for which an information-theoretic reconstruction algorithm was given by Radhakrishnan, R\"{o}tteler and Sen~\cite{RadhakrishnanRS}.
\item $(SL_2(q),B)$ or $(GL_2(q),B)$ where $B$ is the Borel subgroup consisting of upper-triangular matrices.
\item $(S_n,H)$ where $H$ is the \emph{hyperoctahedral group}; this is the centralizer of $(1\,2)(3\,4) \cdots (n-1 \,n)$, or equivalently the wreath product $S_{n/2} \wr \Z_2$, or the symmetry group of the $(n/2)$-dimensional hyperoctahedron.  
\item $(S_n,H)$ where $H=S_m \times S_{n-m}$ for some $0 \leq m \leq n$, i.e., the subgroup of permutations under which the set consisting of the first $m$ elements is invariant.
\end{itemize}
Note there is an efficient classical algorithm for the hidden conjugate problem for the last two examples in $S_n$: simply check for all ${n \choose 2}$ transpositions whether the oracle differs from its value on the identity.  This allows us to determine the conjugate of $H$, which is associated with a matching (for the hyperoctahedral group) or a subset of size $m$ (for $S_m \times S_n$).

\section{Conclusion}

The hidden conjugate problem has important applications; in the dihedral group it is related to hidden shift problems~\cite{vanDamHI03} and lattice problems~\cite{Regev}.  However, for problems such as Graph Isomorphism, we are typically interested in distinguishing one conjugacy class from another.  While we can detect the trivial subgroup with the additional measurement $M_0$ defined here and in~\cite{BaconCvD}, the PGM is not generally optimal in this case [D. Bacon, personal communication].  Constructing the optimal measurement for the hidden subgroup problem, given a prior on the conjugacy classes, remains an important open question.

\section*{Acknowledgments.}  
This work was supported by NSF grants CCR-0093065, PHY-0200909,
EIA-0218443, EIA-0218563, CCR-0220070, and CCR-0220264.  
We are grateful to David Bacon, Andrew Childs, and Wim van Dam 
for introducing us to the Pretty Good Measurement and alerting us to reference~\cite{YuenKM}, 
to the organizers of QIP 2005 at which much of this work was done, to Dan Rockmore and Martin R\"{o}tteler for thoughts on Gel'fand pairs, 
and to Tracy Conrad and Sally Milius for their support and tolerance.  
C.M.\ also thanks Rosemary Moore for her recent arrival, and for providing a 
larger perspective.

\end{document}